# Query Refinement by Multi Word Term expansions and semantic synonymy

Véronika Lux-Pogodalla[a], Eric SanJuan[b]
[a] *SRDI, INIST-CNRS, France*
[b] *LITA, University of Metz -- URI, INIST-CNRS, France*

We developed a system, TermWatch (https://stid-bdd.iut.univ-metz.fr/TermWatch/index.pl), which combines a linguistic extraction of terms, their structuring into a terminological network with a clustering algorithm. In this paper we explore its ability in integrating the most promising aspects of the studies on query refinement: choice of meaningful text units to cluster (domain terms), choice of tight semantic relations with which to cluster terms, structuring of terms in a network enabling abetter perception of domain concepts. We have run this experiment on the 367 645 English abstracts of PASCAL 2005-2006 bibliographic database (http://www.inist.fr) and compared the structured terminological resource automatically build by TermWarch to the English segment of TermScience resource (http://termsciences.inist.fr/) containing 88 211 terms.

## 1 INTRODUCTION

Previous studies have established that information seekers rarely use the enhanced search features available on most search engines or in specialised databases. Average query text consists of 1.8 words [10]. This means that query terms are often too imprecise. Query refinement (QR) aims to improve retrieval performance by adding similar terms to the original user's query. This has the effect of narrowing down the scope of search results or alternatively expanding the search to other related terms with the hope of improving precision. Thus, studies on QR also encompass those on query expansion, query reformulation [9] and relevance feedback [2].

This study specifically addresses the precision improvement problem by defining an automatic way of building a terminological resource and a more linguistic way of refining query terms to their semantic nearest neighbours (SNN). Our methodology favours the extraction of multi-word terms. S-NN terms of a query term are those sharing the same semantic head word (concept) and some common modifier words (properties or qualifiers) as the query term. For instance, the two terms "object software", "object oriented software" are semantically closer to one another than to the third term "objected oriented software testing". While "object oriented software" is an insertion variant (a type) of "object software", the term "object oriented software testing" is a head-expansion of the second term. Head expansions induce a bigger semantic gap between terms, i.e., here, we move from the concept of "software" to that of "testing". This way of identifying S-NN terms does not require document co-occurrence information as it is based on internal linguistic operations between terms. S-NN terms identified in this way are then clustered.

We developed a system, TermWatch (https://stid-bdd.iut.univ-metz.fr/TermWatch/index.pl), which combines a linguistic extraction of terms, their structuring into a terminological network with a clustering algorithm. Contrarily to [4]'s study, the whole process leading from term extraction, term categorisation, clustering and document ranking is automated. It comprises three main modules: a term extractor, relation identifier which yields the terminological network and a clustering module. The results are output on two interfaces: a graphic one mapping the clusters in a 2D space and a terminological hypertext network allowing the user to interactively explore results, return to source texts or re-execute the system's modules.

Since clustering is based on general linguistic relations that are not dependent on a particular domain and it does not do not require specific labour for every text collection. TermWatch (see [13] for a presentation), was designed to achieve the following : choice of meaningful text units to cluster (domain terms), choice of tight semantic relations with which to cluster terms, structuring of terms in a network enabling a better perception of domain concepts.

In this paper we explore the relevancy of these features for query refinement.

The rest of the paper is organised as follows: section 2 presents our methodology; section 3 describes the query refinement experiment and section 4 analyses results; finally section 5 discusses lessons learned from this experiment.

## 2 METHOD

We used TermWatch system to mine noun phrases (NPs) as Multi Word term candidates for a domain thesaurus. This extraction is based on shallow NLP, using the LTPOS tagger and LTChunker software, both developed by the University of Edinburgh. LTChunk identifies simplex noun phrases (NPs), i.e., NPs without prepositional attachments. In order to extract more complex terms, we wrote contextual rules to identify complex terminological NPs.

The system clusters the extracted terms following different combinations of the lexico-semantic relations described above.

A simple and straightforward relation for clustering would be to group together terms sharing the same head (eg. Energy balance and "heat balance"). A more refined clustering is to select a subset of variation relations that induce semantic proximity between terms. Such relations would be spelling variants, substitutions of synonymous words via WordNet (eg. "inflammatory reaction" and "inflammatory response") and left expansions on long terms (eg. "bone marrow cell" and "immature bone marrow cell"). We shall refer to this reduced set of relations as COMP. By computing the set of connected components generated by COMP relations, we obtain small homogeneous clusters. Some of these clusters can have terms with different heads (acquired via WordNet synonyms).

Each component is labelled by its central term, i.e the term having the highest number of variants.

We decided to take this list of components label as terminology resource for query refinement. Whatever are the chosen relations, building progressively these components allow to map user term queries onto a structured network of terminological variants. TermWatch's clusters are output as a interactive hypertext interface that allows the user to navigate among the network stored in a relational database and access documents via the term variants proposed as refined queries.

## 3 EXPERIMENT

We have experimented TermWatch's QR abilities on the 367 645 English abstracts of PASCAL 2005-2006 bibliographic database (http://www.inist.fr) and compared the structured terminological resource automatically build by TermWatch to the English segment of TermSciences resource (http://termsciences.inist.fr/) containing 88 211 terms automatically structured by basic clustering and lexico-semantic relations.

### 3.1 Terminology extraction

We split the corpus into two parts. We used a small part made of the 20 000 first English abstracts of PASCAL 2006 to learn some terminology used in the corpus. The most computer expensive task was the tagging of these 20 000 abstracts by the LTPOS tagger (1h for 3 000 abstracts) meanwhile the extraction of long NP's was immediate. We obtain 308 442 NP's from these 20 000 abstracts.

These NP's were grouped into 75 807 components using variations described in table 1. Thus two terms that are variants of one of these types were grouped together. The resulting components are small connected graphs (less than 30 vertex). The vertex with the highest degree is used as label in TermWatch system. We considered these labels as multiword term (MWT) candidates. This is a way of extracting a clean list of possible MWT without need of any statistics. Indeed, by considering only multiword terms with a relatively high variation activity in the sense of table 1, we reduce the noise due to tagging errors or common NP's like "good paper". Moreover, by avoiding the use of contextual statistics, MWT candidates are selected independently from the way they occur in the small subset of 20 000 abstracts.

Table1. List of COMP variation used to form connected components

| Variation | Description | Number of relations |
|---|---|---|
| **orthograph** | Lexical | 2 844 |
| **exp_l** | Left adjunction of modifiers | 84 999 |
| **ins** | Insertion of modifiers | 4 248 |
| **sub_wn** | Substitution of a word by a synonym | 188 |

We shall now compare the resulting extraction to a reference resource in the Query Refinement perspective.

### 3.2 TermSciences reference

TermSciences [17] is a freely and online accessible terminological database, comprising about 150,000 concepts related to over 540 000 terms, in up to 4 languages: French, German, English and Spanish. It results from a merge of formerly separate databases of terms covering the whole range of scientific domains, including natural and technical sciences (e.g. physics, biology, chemistry, medicine, computer sciences) as well as arts and humanities (e.g. philosophy, linguistics, sociology, education). The terminological data originates in specialized thesauri and vocabularies at INIST, but also in external databases. The main idea behind TermSciences is to make publicly available, for practical applications such as document engineering, knowledge extraction, indexing, but also for research in Natural Language Processing, an important amount of French scientific terms accumulated and maintained over the last 15 years.

Nevertheless, the current version of TermSciences includes heterogeneous data with respect to their initial purpose (manual or automatic indexing, translation, etc.) and, consequently, to their surface form (order of constituents, inflected forms, abbreviations, etc.). Work is ongoing to improve the quality of TermSciences and to make the resource more homogeneous.

Here, we use a subset of the current TermSciences : 88 211 English terms among which 58 683 are multiword terms from scientific domains covered by PASCAL database. These terms were used to index bibliographical notices, initially manually. A notice contained in PASCAL is a document with different sections, in particular : title, authors, keywords, abstract.

## 4 RESULTS

We first looked for the occurrence of MWT from each resource TermWatch (TW) and TermSciences (TS) in PASCAL 2005 – 2006. Table 2 compares their frequencies.

Table2. Frequence of terms by resource in PASCAL 2005-2006

|  | *Number of MWT* | *MWT matching abstracts* | | *Number of abstracts* | | *Maximal frequency* |
|---|---|---|---|---|---|---|
|  |  | Length>1 | Length>2 | Length>1 | Length>2 |  |
| ***TermWatch*** | 75 807 | 34 060 | 5 303 | 81 140 | 9 441 | 4 279 |
| ***TermSciences*** | 58 683 | 3 673 | 12 | 18 169 | 339 | 1 942 |

A very striking result is the little number of occurrences of TermSciences terms in the abstracts. This should be related to the fact that TermSciences (or more precisely, the largest vocabularies included in TermSciences) is a mainly manually built resource (vs. a corpus based resource like the one extracted by TermWatch) that was mainly built for indexing and, at least initially, for manual indexing.
TermSciences therefore includes many terms that are not in natural language but in an artificial language in which the order of constituents may be different (eg. "Dementia, Alzheimer Type" and "Keynes (J.M)" are terms.). Such terms were suitable for paper index.
TermSciences also includes terms that are unlikely to occur in text as such and look more like a "label on a concept" (eg. "masculine-feminine" is a term and labels a concept that can appear in text as "… tracing the sexual difference …" or "Overcoming the hierarchical dichotomy of male and female …",)
It also includes terms in which preposition and determiners were deleted (eg. the French term "moule coulée précision" is registered. It probably occurs in texts as "moule pour coulée de précision").
Furthermore, English terms in TermSciences can be translation from the French terms – and translators had no corpus to support their work (eg. "clean and unclean" seems a translation from the French "Pur-Impur". Abstracts of notices indexed by "clean and unclean" for includes sentences such as "Here the profane is the impure; the sacred, the pure.")
These results clearly show the extend to which a hand-made controlled vocabulary used for (manually)

indexing is different from a corpus-based terminology. The difference is huge indeed !

As a consequence, such vocabulary is not adequate as such for text mining/querying. So the next question is : how can we use TermWatch to refine queries made with the TermSciences vocabularies ?

To answer this question, we compared the two resources, considering TermWatch label components as possible refinements of TermSciences terms : as the following table shows, 5 070 TermSciences terms have a left right expansion (LR-exp) in TermWatch (the TS term has to appear as a substring of at least one TW term).

Table3. Number of terms in TW and TS related by left right expansion (LR-exp)

|  | **Uniterms** | **MWT** |
|---|---|---|
| *Number of TW terms that are LR-exp of TS terms* | 0 | 8 532 |
| *Number of TS terms that have a LR-exp in TW* | 4 667 | 5 070 |

Let us get an insight view into the possible expansions of TermSciences terms that can be founded in TermWatch. Table 4 gives the proportion of MWT expansions founded per number of added words.

Table4. Repartition of TW expansions on the number of added words

| *Difference of length between TW and TS terms* | 0 | 1 | 2 | 3 |
|---|---|---|---|---|
| *Proportion of TW terms* | 0.53 | 0.62 | 0.3 | 0.07 |
| *Proportion of TS terms* | 0.8 | 0.75 | 0.34 | 0.1 |

In Table 4, we can see that among the 5 070 TermSciences terms included in at least one TermWatch candidate term, there are more exact matchs (80%) than one word expansions (75%). This suggests that terms of an artificial indexing vocabulary are not adequate starting terms for trivial LR-expansions (substring occurrence). Taking into account other types of relations (like insertions and WordNet substitutions from table 1), TermSciences terms can be related to many more TermWatch terms. These terms are likely to be relevant in QR perspective because, as showed in [13], they belong to clusters that are semantically homogeneous.

Table5. Number of TW terms that are variants of TS MWT.

| *Number of variations to be applied* | *0* | *1* | *2* | *3* |
|---|---|---|---|---|
| *TW terms* | 5 070 | 15 640 | 19 301 | 20 426 |

Last, we observed that TermSciences uniterms seem to be much "too generic" to be considered as queries. This is because TermSciences vocabulary was meant to be used in a "post-coordinated" manner when used for searching. TermWatch is a useful resource here to show which combinations of uniterms really occur in corpora. As table 6 shows, a significant number of TermWatch MWT candidate terms (ie. 19 198) include several TermSciences uniterms and the total number of uniterms involved in TermWatch candidates by this way is 4 668.

Table6. Number of TW terms that include several TS uniterms.

| *Number of uniterms* | *2* | *3* | *4* | *>4* |
|---|---|---|---|---|
| *Proportion TW terms* | *14 103* | *4 320* | *680* | *95* |

Here are some examples of such TW terms: *city traffic signal datum acquisition system* that contains the TS uniterms *acquisition, signal, system and traffic*; *oral tumor cell proliferation activity influence* that contains TS uniterms *activity, cell, influence, proliferation.*

## 5 DISCUSSION

Our initial goal was to evaluate TermWatch possibilities in QR considering terms of a reference vocabulary, here TermSciences, as initial requests.

Major lessons learned from the experiments are about the differences between a manually built controlled vocabulary, which TermSciences still largely is, and a corpus-based corpus-based set of candidates terms (eg. TermWatch).
But despite this feature, the experiments demonstrate TermWatch abilities in QR. In particular, we have shown that, using a larger set of relations than initially planed, TermWatch can refine a significative proportion of requests made with TermSciences terms and that many uniterms of TermSciences can also be refined as multi-words queries.

A key point to improve the QR given the TermSciences peculiar characteristics now lies in our ability to model as additional particular variations the relations between terms extracted from a corpus and their correspondents in the controlled vocabulary. This is our next challenge.